\documentclass[journal]{IEEEtran}

\usepackage[T1]{fontenc}
\usepackage{graphicx,latexsym,subfigure,amsmath,epsfig,latexsym,subfigure,amsmath,cite,amssymb}
\usepackage{algorithm,amsthm,epstopdf}
\usepackage{algorithmic}
\usepackage{color}
\usepackage{times}
\usepackage{subeqnarray}
\usepackage{cases}
\usepackage{setspace}
\usepackage{url}
\usepackage{diagbox}
\usepackage{bm}
\usepackage{slashbox,multirow}
\usepackage{indentfirst}
\usepackage{CJK}
\usepackage{stfloats}
    \newtheorem{theorem}{Theorem}

\begin{document}
\title{Learning-Based Computation Offloading for IoT Devices with Energy Harvesting}
\author{Minghui Min\IEEEauthorrefmark{1}, Dongjin Xu\IEEEauthorrefmark{1}, Liang Xiao\IEEEauthorrefmark{1}, Yuliang Tang\IEEEauthorrefmark{1}, Di Wu\IEEEauthorrefmark{2}\\
\IEEEauthorblockA{\IEEEauthorrefmark{1}Dept. of Communication Engineering, Xiamen University, Xiamen, China. Email: \{lxiao, tyl\}@xmu.edu.cn}\\
\IEEEauthorblockA{\IEEEauthorrefmark{2}Dept. of Computer Science, Sun Yat-Sen University, Guangzhou, China. Email: wudi27@mail.sysu.edu.cn}}
\maketitle

\begin{abstract}
  Internet of Things (IoT) devices can apply mobile-edge computing (MEC) and energy harvesting (EH) to provide the satisfactory quality of experiences for computation intensive applications and prolong the battery lifetime. In this article, we investigate the computation offloading for IoT devices with energy harvesting in wireless networks with multiple MEC devices such as base stations and access points, each with different computation resource and radio communication capability. We propose a reinforcement learning based computation offloading framework for an IoT device to choose the MEC device and determine the offloading rate according to the current battery level, the previous radio bandwidth to each MEC device and the predicted amount of the harvested energy. A ``hotbooting'' Q-learning based computation offloading scheme is proposed for an IoT device to achieve the optimal offloading performance without being aware of the MEC model, the energy consumption and computation latency model. We also propose a fast deep Q-network (DQN) based offloading scheme, which combines the deep learning and hotbooting techniques to accelerate the learning speed of Q-learning. We show that the proposed schemes can achieve the optimal offloading policy after sufficiently long learning time and provide their performance bounds under two typical MEC scenarios. Simulations are performed for IoT devices that use wireless power transfer to capture the ambient radio-frequency signals to charge the IoT batteries. Simulation results show that the fast DQN-based offloading scheme reduces the energy consumption, decreases the computation delay and the task drop ratio, and increases the utility of the IoT device in dynamic MEC, compared with the benchmark Q-learning based offloading.
\end{abstract}

\section*{Introduction}
With limited computation, power and memory supplies, Internet of Things (IoT) devices such as sensors, cameras and wearable devices have the computation bottleneck to support the advanced applications such as interactive online gaming and face recognition \cite{mao2016dynamic}. This challenge can be addressed by the mobile edge computing (MEC) techniques, in which IoT devices offload the computation tasks to the MEC devices such as the base stations, access points (APs), laptops and even smartphones within the radio access of the IoT devices. By utilizing the computation, cache and power resources of the MEC devices, computation offloading can reduce the computation delay, save the battery lifetime, and even enhance security for the computation intensive IoT applications \cite{mao2017survey}. Energy harvesting (EH) is another promising technique to prolong the battery lifetime and provide satisfactory quality of experiences for IoT devices \cite{ulukus2015energy}. Equipped with EH modules, an IoT device can capture the ambient renewable energy, such as the solar radiation, wind and human motion, and other resources, such as the ambient radio-frequency (RF) signals, to supply the ``greener'' energy for the central processing unit (CPU) and the radio transceiver.

Resource allocation and partial offloading in a multiuser MEC network are investigated in \cite{You7762913} to minimize the energy consumption under the latency constraint, in which each user is powered by fixed energy sources such as battery. The binary offloading policy in a wireless powered MEC network is studied in \cite{bi2017computation}, in which the AP is used to transfer RF power to the mobile device and  execute the computation tasks. The multiple-antenna AP to power the users is studied in \cite{wang2017joint} to minimize the AP's total energy consumption subject to the users' individual latency constraints.

A critical problem in the computation offloading is to choose the MEC device from all the radio device candidates in the radio coverage and determine the offloading rate, that is, the amount of the computation tasks to offload to the MEC device. However, an IoT device sometimes takes even longer time to send the offloading data and receive the computation result compared with the local computation on the IoT device, if the chosen MEC device is carrying out heavy workloads and experiencing degraded radio channel fading and interference \cite{xu2017online}. It is challenging for an IoT device to optimize the offloading policy in the dynamic MEC network with time variant radio bandwidths, and unknown amount of the harvested energy in a given time duration.

The offloading issue has recently attracted research attentions. For example, the mobile offloading scheme as proposed in \cite{mao2016dynamic} uses the Lyapunov optimization to minimize the worst-case expected computation cost with a single known MEC server, assuming the knowledge of both the transmission delay model and the local execution model. The online learning based mobile edge computing with energy harvesting in \cite{xu2017online} can reduce the time cost assuming the normal distributed renewable energy generation under the known offloading delay model and the power consumption model. However, these assumptions do not always hold for practical IoT devices, especially in dynamic MEC with multiple MEC candidates.

In this article, we investigate the computation offloading of IoT devices with energy harvesting in dynamic MEC. Without loss of generality, this work assumes that the IoT device is connected with multiple MEC devices of different computation and communication overheads, and can use the EH history to approximately predict the future renewable energy generation trend. We present a reinforcement learning (RL) based computation offloading framework for an IoT device to choose the MEC device and determine the proportion of the computation tasks to offload. Each offloading decision is made according to the radio bandwidth for each IoT-MEC link in the last time slot, the predicted amount of harvested energy and the current battery level of the IoT device, which are defined as the system state in the repeated game with the MEC network. As we will see, the computation offloading process of the IoT device can be viewed as a Markov decision process (MDP). Consequently, RL techniques such as Q-learning enables the IoT device as a learning agent to achieve the optimal offloading strategy after a large number of offloading interactions with probability 1 \cite{Qintroduction}.

We propose a ``hotbooting'' Q-learning based computation offloading scheme for an IoT device with EH without being aware of the MEC model, the energy consumption and computation latency model. The offloading strategy including the MEC device and the offloading rate is chosen according to the current MEC state and the quality function or Q-function that is the expected long-term discount reward for each state-action pair. The IoT device evaluates the reward or utility based on the overall delay, energy consumption, the task drop loss and the data sharing gains in each time slot, and updates the Q-function according to the iterative Bellman equation based on the utility and the offloading strategy. As a transfer learning method in reinforcement learning \cite{pan2010survey}, the hotbooting technique exploits the offloading experiences in similar scenarios to initialize the Q-function and thus to save the exploration time at the initial stage in the repeated offloading game with MEC.

We also propose a fast deep Q-network (DQN) based offloading scheme to further improve the offloading performance for the case with a large state space, for example, a large number of feasible IoT battery levels, a large number of MEC devices, a large amount of the renewable energy generated in a time slot and a large number of potential radio bandwidths to each MEC device. By applying the deep reinforcement learning technique to compress the state space in the Q-learning process, this scheme uses the deep convolutional neural network (CNN) to estimate the Q-value for each action \cite{mnih2015human}, and thus accelerate the learning speed for the IoT device that has sufficient computational resources to implement deep learning.

We analyze the performance of the RL-based offloading schemes and provide the convergence performance for two typical MEC scenarios, assuming an energy consumption model based on the CPU frequencies and the transmit power. We consider the local computational time and  the transmission time in the analysis and provide the performance bound of the proposed scheme regarding the utility of the IoT from each offloading decision. As a concrete example, simulations are performed for IoT devices with wireless power transfer (WPT) to capture the ambient RF signals from a dedicated RF energy transmitter to charge the IoT battery \cite{Zhang7081084}. Simulation results show that the fast DQN-based offloading scheme reduces the energy consumption of IoT devices, decreases the computation delay and the task drop ratio, and increases the utility of IoT devices, compared with the benchmark Q-learning based offloading scheme. The main contributions of this paper are summarized as follows:

\begin{itemize}
\item We present a hotbooting Q-learning based computation offloading scheme for IoT devices with energy harvesting to achieve the optimal offloading without knowing the MEC model, the energy consumption and computation latency model in dynamic MEC with low computation complexity.

\item We also propose a fast DQN-based IoT computation offloading scheme to compress the state space dimension and a hotbooting technique to accelerate the learning speed at the initial stages and improve the offloading performance for IoT devices.

\item We provide the performance bound of the proposed RL-based offloading schemes and prove their convergence to the optimal utility.
\end{itemize}

The rest of this paper is organized as follows. We first present the mobile offloading model and the energy harvesting model. Then we propose a hotbooting Q-learning based mobile offloading scheme for IoT devices with energy harvesting and a fast DQN-based offloading scheme. Following that, we analyze the performance bounds of the proposed mobile offloading schemes and present simulation results. Finally, we conclude the article.

\section*{System Model}
\begin{figure}[!htbp]
\begin{center}
\includegraphics[height=2.2 in]{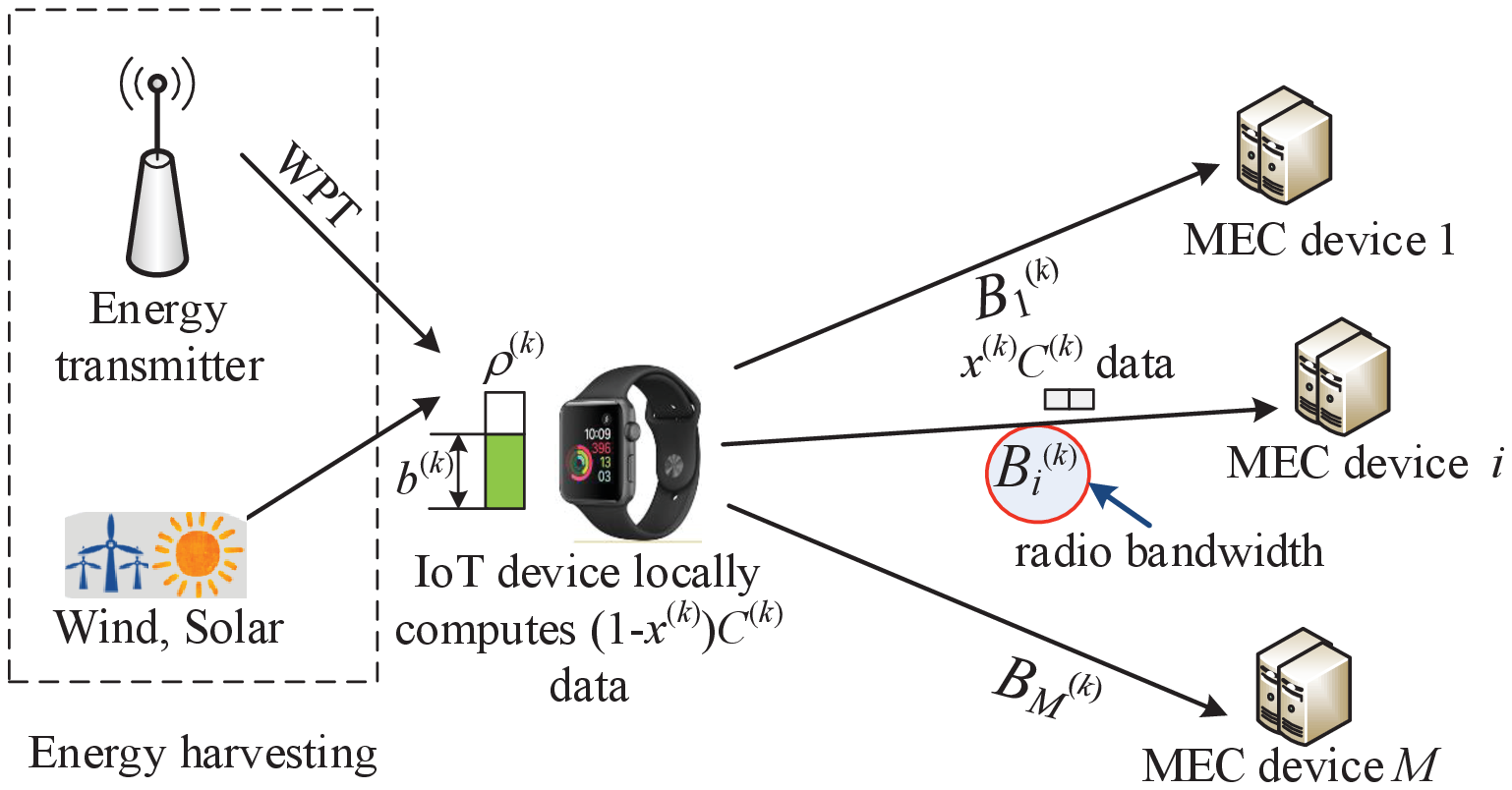}\\
\caption{Mobile-edge computing system with an energy harvesting IoT device and $M$ MEC devices.}\label{system1}
\end{center}
\end{figure}

We consider an MEC network consisting of an IoT device and $M$ MEC devices as shown in Fig. \ref {system1}. The IoT device is equipped with electricity storage devices and an EH component, such as the wind turbines, photovoltaic modules and RF energy harvester. With the energy powered by the EH component, the IoT device can execute its computation task locally and offload the task to the MEC devices that can provide the high computation performance with a virtual machine. We assume that time is slotted, and denote the time slot index by $k$, with $k \in \mathcal{K}\doteq \{0,1,...\}$.

The size of the computation task generated by the IoT device at time slot $k$ is denoted by $C^{(k)}$ (in bits). Similar to \cite{ mao2016dynamic}, we focus on delay-sensitive applications with the execution duration of the computing task $C^{(k)}$ no greater than the time slot length. The IoT device chooses MEC device $i$ to offload the computation task based on the radio channel state, with $1\leq i\leq M$, and the proportion of the data offloaded to MEC device $i$ denoted by $x^{(k)}$, with $0 \leq x^{(k)} \leq 1$, to improve the computation performance. More specifically, the IoT device executes the computation tasks locally if $x^{(k)} = 0$, and offloads all the computation tasks to MEC device $i$ if $x^{(k)} = 1$. The IoT device offloads $x^{(k)}C^{(k)}$ data to the MEC device and the rest $(1-x^{(k)})C^{(k)}$ data are processed locally if $0<x^{(k)}<1$. For simplicity, the offloading rate $x^{(k)}$ is quantized into $N_x + 1$ levels, that is, $x^{(k)} \in \{l/N_x\}_{0 \leq l \leq N_x}$. The IoT device chooses its action at time slot $k$ denoted by $\textbf{a}^{(k)} = [i,x]\in \textbf{A}$ from the vector space of all the possible actions denoted by $\textbf{A}$.
\subsection* {Computation-Offloading Model}
The IoT device sends the computation tasks to the MEC device $i$ at time slot $k$ with the radio bandwidth $B_i^{(k)}$. The transmission duration to offload $x^{(k)}C^{(k)}$ bits data to MEC device $i$ is represented by $T_i^{(k)}$, with $T_i^{(k)}={x^{(k)}C^{(k)}}/{B_i^{(k)}}$.

According to \cite{Chen7307234}, the IoT device consumes $E_{i}^{(k)}$ energy to offload the tasks to MEC $i$ at time slot $k$, with $E_{i}^{(k)}= {P^{(k)} x^{(k)} C^{(k)}}/{B_i^{(k)}}$, which depends on the transmit power for offloading $P^{(k)}$ and the transmission duration $T_i^{(k)}$.
\subsection* {Local-Computing Model}
The CPU of an IoT device is the primary engine for local computation. The CPU performance is controlled by the CPU-cycle frequency. Consider the local computing for executing $(1-x^{(k)})C^{(k)}$ bits of input data at the IoT device. Let $N$ denote the number of CPU cycles required for computing one input bit. The total number of CPU cycles required for the $(1-x^{(k)}) C^{(k)}$ bits is $(1-x^{(k)})C^{(k)}N$. By applying dynamic voltage and frequency scaling techniques \cite{mao2017survey}, the IoT device can control the energy consumption for local task execution by adjusting the CPU frequency $f_m$. In practice, $f_m$ is bounded by a maximum value $f^{max}$, indicating its computation capability limitation. The execution latency for local computing at time slot $k$ is denoted by $T_{0}^{(k)} $, with $T_{0}^{(k)} = \sum _{m=1}^{(1-x^{(k)})C^{(k)}N} {1}/{f_m}.$

According to \cite{zhang2013energy}, the IoT device consumes $E_{0}^{(k)}$ energy for local computing at time slot $k$, with $E_{0}^{(k)} =  \sum _{m=1}^{(1-x^{(k)})C^{(k)}N} \varsigma f_m^2$, where $\varsigma$ is the effective capacitance coefficient that depends on the chip architecture.
\subsection* {Energy Harvesting Model}
IoT devices can be charged by renewable resources, such as solar and wind, and other resources, such as the ambient RF signals. The energy harvested at time slot $k$ denoted by $\rho^{(k)}$ is stored in a battery of the IoT device to support both the local execution and offloading, and $\rho^{(k)}$ can be estimated by the IoT device according to the energy harvesting history and the modeling method such as \cite{wang2016incentivizing}. The estimated amount of the harvested energy at time slot $k$ is denoted by $\hat{\rho}^{(k)}$. The estimation error regarding $\rho^{(k)}$ is denoted by $\bigtriangleup ^{(k)}$, and $\bigtriangleup ^{(k)} = \rho^{(k)}- \hat{\rho}^{(k)}.$



The battery level at the beginning of time slot $k$ denoted by $b^{(k)}$, and it evolves according to the following equation:
\begin{align}
b^{(k+1)} = b^{(k)}-\left(E_0^{(k)}+ E_i^{(k)}\right)+\rho^{(k)},
\end{align}
in which $(E_0^{(k)}+ E_i^{(k)})$ represents the total energy consumed by the IoT device at time slot $k$.
Note that the IoT device drops the computation task, if the energy is insufficient at the IoT device (i.e., $b^{(k+1)}<0$) \cite{mao2016dynamic}.

As a special case, we consider a wireless powered MEC model as shown in Fig. \ref {system1}, in which the energy transmitter broadcasts wireless energy to charge the IoT device. Let $\eta^{(k)}$ define the transmit power of the energy transmitter and $h^{(k)}$ represent the wireless channel gain between the energy transmitter and the IoT device at time slot $k$, and $\upsilon \in (0,1)$ denote the energy harvesting efficiency. According to \cite{Zhang7081084}, the amount of the harvested energy of the IoT device in one unit of time slot is denoted by $\tilde{\rho}^{(k)}$, with $\tilde{\rho}^{(k)} = \upsilon \eta^{(k)} h^{(k)}$.

With EH IoT devices, the computation offloading strategies dedicated for MEC system become much more complicated compared to that of conventional battery-powered devices. Specially, both the energy state information and the channel state information need to be considered, and the temporally correlated battery energy level links the system decisions in different time slots. Consequently, it brings new challenge to derive an optimal computation offloading strategy to achieve satisfactory quality of experiences for users.
\section*{Hotbooting Q-learning Based Computation Offloading}
In the dynamic computation offloading process, the IoT device chooses its
proportion of data to offload to an MEC device based on the system state, which consists of the previous radio bandwidth to each MEC device, the predicted amount of harvested energy and the current battery level. The next system state observed
by the IoT device is independent of the previous states and actions,
for a given system state and computation offloading strategy in the current
time slot. Therefore, the computation offloading process can be viewed as an MDP, in which the Q-learning technique, as a model-free and widely used RL technique, can derive the optimal policy without being aware of the MEC model, the energy consumption and computation latency model.

We propose a hotbooting Q-learning based offloading scheme in the dynamic offloading process, in which the hotbooting technique is used to initialize the Q-value with the computation offloading experiences in similar environments\cite{lu2017}. The hotbooting Q-learning based offloading scheme saves the random explorations at the beginning stage of the dynamic computation offloading process and thus accelerates the learning speed.
The Q-values as the output of the hotbooting process is used to initialize the Q-values at the beginning of the learning process.

At time slot $k$, the IoT device observes the system state of the MEC system, $\textbf{s}^{(k)}$, which consists of the previous radio bandwidth, the predicted amount of the harvested energy and the current battery level, that is, $\textbf{s}^{(k)} = \left[B_1^{(k-1)},..., B_M^{(k-1)}, \hat{\rho}^{(k)}, b^{(k)}\right]$.

Let $Q(\textbf{s},\textbf{a})$ denote the Q-function. Based on the state $\textbf{s}^{(k)}$, the IoT device uses $\varepsilon$-greedy policy to choose the computation offloading action $\textbf{a}^{(k)}\in \textbf{A}$ \cite{Qintroduction}, including the selected $i$-th MEC device and the offloading rate $x \in [0,1]$.
More specifically, the offloading policy that maximizes the Q-function is chosen with a high probability $1-\varepsilon$, while other offloading actions are selected with an equal low probability $\varepsilon/\left(\left(N_x+1\right) M\right)$ to avoid staying in the local maximum.

The IoT device fails to perform the computation task if the energy is insufficient at the IoT device (i.e., $b^{(k+1)}<0$). The task drop loss denoted by $\psi$ is defined as the cost if a computation task fails. The indicator function denoted by $\mathbf{I}\left(\varpi\right)$ equals 1 if $\varpi$ is true and 0 otherwise. Let $\beta$ and $\mu$ be the weighting parameters of energy consumption and the computation delay of the IoT device, respectively.
The computation delay of the IoT device at time slot $k$, $\max \{T_0^{(k)}, T_i^{(k)}\}$, depends on the local execution latency $T_0^{(k)}$ and the transmission delay to the chosen MEC device $T_i^{(k)}$. The utility of the IoT device to choose the MEC $i$ at time slot $k$ is denoted by $U_{i}^{(k)}(x)$, which depends on the overall data sharing gains, the task drop loss, the energy consumption and the computation delay is given by
\begin{align}\label{U}
U_{i}^{(k)}(x)&= x C^{(k)}- \psi \mathbf{I}\left(b^{(k+1)} \leq 0\right) \cr &-\beta \left(E_0^{(k)}+E_{i}^{(k)}\right) - \mu \max \left\{T_0^{(k)},T_{i}^{(k)}\right\}.
\end{align}

The Q-function is updated according to the iterative Bellman equation based on the utility and the offloading policy as given in \cite{Qintroduction}. The hotbooting Q-learning based offloading scheme can achieve the optimal computation offloading strategy via trial-and-error over sufficient time slots. The learning time of the Q-learning scheme increases with the dimension of the action-state space, which increases with the number of feasible IoT battery levels, the amount of the renewable energy generated in a time slot and the number of potential radio bandwidths to each MEC device, yielding serious performance degradation.

\begin{figure*}[!htbp]
\begin{center}
\includegraphics[height= 3.0 in]{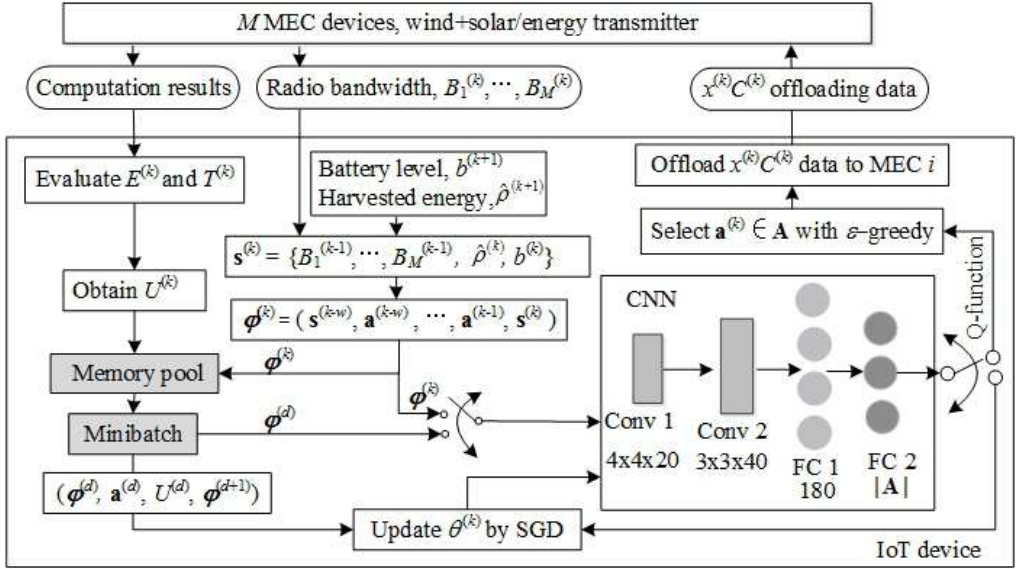} \\
\caption{DQN-based computation offloading scheme.}\label{DQN-based system}
\end{center}
\end{figure*}
\section*{Fast DQN-Based Computation Offloading}
In this section, we propose a fast DQN-based computation offloading scheme to improve the computation performance. As shown in Fig. \ref{DQN-based system}, this scheme utilizes CNN as nonlinear approximator of the Q-value for each action, which allows a compact storage of the learned information between similar states to further accelerate the convergence rate of the hotbooting Q-learning scheme.

The system parameters $C$, $f$ and $N$ are initialized at the beginning based on the nature of application, and $B_1^{(0)},..., B_M^{(0)}$, $\hat{\rho}^{(1)}$ and $b^{(1)}$ are chosen randomly from the corresponding action set. The experience size in the CNN input sequence $W$ and the size of minibatch of the CNN $H$ are initialized to achieve a good computation offloading performance. Besides, the experience pool is set as $\mathcal{D} = \emptyset$, and the CNN weights $ \theta$ is initialized by using the hotbooting technique.

The fast DQN-based offloading scheme as summarized in Algorithm \ref {DQN} extends the system state $\textbf{s}^{(k)}$ as in the hotbooting Q-learning scheme to the experience sequence at time slot $k$ denoted by $\boldsymbol{\varphi}^{(k)}$ to accelerate the learning speed, and improve the computation performance of the IoT device. More specifically, the experience sequence consists of the current system state $\textbf{s}^{(k)}$ and the previous $W$ system state-action pairs, that is,
$\boldsymbol{\varphi}^{(k)} = \left(\textbf{s}^{(k-W)}, \textbf{a}^{(k-W)}, ...,\textbf{a}^{(k-1)}, \textbf{s}^{(k)} \right)$.

If the number of offloading experiences is less than $W$ (i.e., $k\leq W$), the IoT device chooses the offloading policy randomly from the action set $\textbf{A}$.

\renewcommand{\baselinestretch}{1.0}
\begin{algorithm}[!t]
\centering
\begin{algorithmic}[1]
\STATE Initialize the experience size in the CNN input sequence $W$, the size of minibatch of the CNN $H $, and the memory pool $\mathcal{D}$
\STATE Hotbooting process for initializing the CNN weights $\theta$
\FOR {$k = 1, 2, 3,...$}
\STATE Observe current system state $\textbf{s}^{(k)}$
\IF {$k\leq W$}
    \STATE Choose the MEC device $i$ and offloading rate $x^{(k)}$ randomly from the action set $\textbf{A}$
    \STATE Send $x^{(k)} C^{(k)}$ data to the selected MEC $i$
\ELSE
\STATE Observe the experience sequence $\boldsymbol{\varphi}^{(k)}$
\STATE Set $\boldsymbol{\varphi}^{(k)}$ as the input of the CNN
\STATE Observe the output of the CNN to obatain the Q-values $Q\left(\boldsymbol{\varphi}^{(k)}, \textbf{a}\right)$
\STATE Select the MEC device $i$ and offloading rate $x^{(k)}$  via $\varepsilon$-greedy policy \cite{Qintroduction}
\ENDIF
\STATE Observe the radio bandwidth to the $M$ MEC devices $B_1^{(k)},..., B_M^{(k)}$ and the battery level $b^{(k+1)}$
\STATE Estimate the amount of the harvested energy $\hat{\rho}^{(k+1)}$
\STATE Evaluate the energy consumption, the computation delay and the task drop loss
\STATE Obtain the utility $U^{(k)}$ via (\ref {U})
\STATE Observe $\boldsymbol{\varphi}^{(k+1)}$
\STATE Store the new experience $\textbf{e}^{(k)}$ in the memory pool $\mathcal{D}$
\STATE Select experiences from $\mathcal{D}$ at random for $H$ times
\STATE Update the CNN weights with $\theta^{(k)}$ via SGD
\ENDFOR
\end{algorithmic}
\caption{Fast DQN-based computation offloading scheme}\label{DQN}
\end{algorithm}

As shown in Fig. \ref{DQN-based system}, the experience sequence $\boldsymbol{\varphi}^{(k)}$ is reshaped into a square matrix and then input into the CNN. The output of the CNN provides the Q-values (i.e., $Q(\boldsymbol{\varphi}^{(k)}, \textbf{a})$) for all $|\textbf{A}|$ computation offloading policies at the current system sequence $\boldsymbol{\varphi}^{(k)}$. To make a tradeoff between exploitation and exploration, the IoT device selects the offloading action $\textbf{a}^{(k)} \in \textbf{A}$ based on the output of the CNN according to the $\varepsilon$-greedy policy \cite{Qintroduction}, and then sends $x^{(k)}C^{(k)}$ data to the MEC device $i$.

The IoT device evaluates its reward or utility $U^{(k)}$ based on the overall delay, energy consumption, the task drop loss and the data sharing gains in time slot $k$. The DQN-based computation offloading scheme also updates a quality function $Q$ for each action-state pair in the dynamic computation offloading process.

The IoT device applies a experience replay technique in which it stores the offloading experiences at each time slot, $\textbf{e}^{(k)} = \left(\boldsymbol{\varphi}^{(k)}, \textbf{a}^{(k)}, U^{(k)}, \boldsymbol{\varphi}^{(k+1)} \right)$, in a memory pool $\mathcal{D} = \left\{\textbf{e}^{(1)}, ..., \textbf{e}^{(k)}\right\}$. The IoT device randomly chooses experiences from $\mathcal{D}$ and uses stochastic gradient descent (SGD) algorithm to update the CNN weights $\theta^{(k)}$ \cite{mnih2015human}.

%

Similar to the hotbooting Q-learning scheme, the hotbooting technique is applied to initialize the CNN parameters in the fast DQN-based computation offloading rather than initialize them randomly to accelerate the learning speed. The IoT device stores emulational experience $\{\boldsymbol{\varphi}^{(k)}, \textbf{a}^{(k)}, U^{(k)}, \boldsymbol{\varphi}^{(k+1)}\}$ in the database, and performs minibatch update. The resulting CNN weights are used to perform the initialization in Algorithm \ref{DQN}.

\section*{Performance Evaluation}
The optimal offloading policy is analyzed for a single time slot, which can be used as the performance upper bound to evaluate the RL-based offloading scheme in the dynamic process over sufficient time slots. The time index $k$ is omitted in the superscript if no confusion occurs. The IoT device chooses the optimal offloading policy $\textbf{a}^{*}=\left[i^*, x^*\right]$ with $x^*$ of computation tasks offloaded to the MEC device $i^*$ to maximize the utility of the IoT device. According to \cite{mao2016dynamic} and \cite{wang2017joint}, we let the local CPU-cycle frequencies be the same for simplicity (i.e., $f_m=f, 1\leq m\leq (1-x)CN$), to achieve the optimal frequencies of the $N$ CPU cycles scheduled for a single computation task.

\begin{theorem}\label{t111}
If $\max\left\{(\mu+\beta P)/(\beta\varsigma Nf^2+1), 1/(\varsigma  Nf^2)\right\} \leq B_{i^*}$, where $i^*=\arg \max_{1\leq i\leq M}B_i$, the RL-based offloading scheme can achieve the optimal offloading strategy $\textbf{a}^{*}=\left[i^*, 1\right]$ with probability 1, after a sufficiently long time to offload. The utility of the IoT device after convergence is given by
\begin{align}\label{op1}
U= C\frac{ B_{i^*}-\beta P-\mu}{B_{i^*}}- \psi \mathbf{I}\left(b- \frac{P C}{B_{i^*}}+ \rho \leq 0\right).
\end{align}
\end{theorem}

According to \cite{Qintroduction}, Q-learning can eventually derive the optimal offloading policy of the IoT device in the MDP process. If the transmission delay and energy consumption to process the computation tasks are low, the optimal offloading policy of the IoT device is $\textbf{a}^{*}=\left[i^*, 1\right]$, indicating that the IoT device offloads all the computation tasks to the MEC device with the largest radio bandwidth to maximize its utility given by (\ref {op1}).

\begin{theorem}\label{t112}
If $B_{i^*}\left(f+\beta\varsigma  N f^3 +\mu N \right)\leq \beta P f$, where $ i^*=\arg \max_{1\leq i\leq M}B_i$, the RL-based offloading scheme can achieve the optimal offloading strategy $\textbf{a}^*=[i^*,0]$ with probability 1, after a sufficiently long time to offload. The corresponding utility of the IoT device after convergence is given by
$U= - \psi \mathbf{I}\left(b-\varsigma CN f^2 + \rho \leq 0\right) - CN\left(\beta \varsigma f^2+  \mu /f\right)$.
\end{theorem}

If the transmission delay between the IoT device and the MEC device with the largest bandwidth is high, the IoT device processes all the computation tasks locally to maximize its utility as shown in Theorem \ref {t112}.

\section*{Simulation Results}
Simulations have been performed to evaluate the performance of the proposed RL-based computation offloading schemes for IoT devices with WPT to charge the battery in dynamic networks with 3 MEC devices. Let the estimation error follow a uniform distribution with the maximum estimation error $G$ (i.e., $\bigtriangleup ^{(k)}\sim G\cdot U(-1,1)$). The experience size in the CNN input sequence $W$, the size of minibatch of the CNN $H$, $\alpha$, $\gamma$ and $\varepsilon $ are initialized to achieve a good computation offloading performance according to the experiments not presented in this article.
For sake of comparison, we compared the proposed schemes with the benchmark Q-learning.

\begin{figure}[!t]
\begin{center}
\includegraphics[height= 2.0 in]{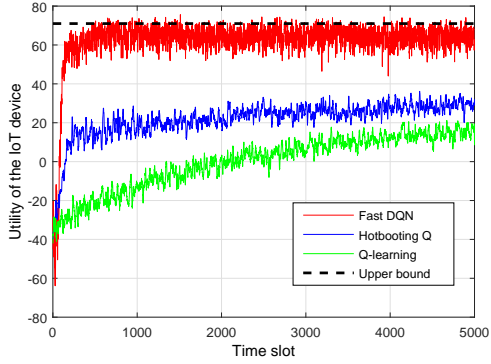} \\
\caption{Performance of the IoT computation offloading over time, in which the MEC network has an IoT device and three MEC devices, with the size of the computation task $C = 100$ bits, the transmit power of the energy transmitter 6 $W$, and $B_1 \in \{3, 4, 5, 6, 7\}$, $B_2 \in \{11, 12, 13, 14, 15\}$ and $B_3 \in \{6, 7, 8, 9, 10\}$.} \label {Dynamicperformance1}
\end{center}
\end{figure}

Fig. \ref {Dynamicperformance1} shows the dynamic learning curves of the RL-based computation offloading schemes over time and how they converge to the performance bound. The IoT device generates 100 bits of computation task in a time slot, the transmit power of the energy transmitter is 6 $W$, and the radio bandwidths of the links from the IoT device to the three MEC devices are randomly chosen with $B_1 \in \{3, 4, 5, 6, 7\}$, $B_2 \in \{11, 12, 13, 14, 15\}$, and $B_3 \in \{6, 7, 8, 9, 10\}$. As shown in  Fig. \ref {Dynamicperformance1}, the fast DQN-based offloading strategy achieves the optimal computation offloading performance after convergence, which matches the analysis result in Theorem \ref {t111}. For example, the utility of the IoT device almost converges to the performance bound given by (\ref {op1}), if the simulation settings satisfy the conditions in Theorem \ref{t111}. In this case, the IoT device offloads all the computation tasks to the MEC device to maximize the system performance.

We can also see that the hotbooting Q-learning based computation offloading exceeds the Q-learning offloading scheme, by having a faster learning speed and a higher utility. The computation offloading with fast DQN further improves the performance, for example, it increases the utility of the MEC network by 2 times at time slot 1000, compared with the hotbooting Q-learning scheme. The fast DQN-based scheme converges within approximately 1000 time slot, while the other two schemes need much more time slots to converge. That's because the fast DQN-based algorithm, an extension of Q-learning, compresses the learning state space by using CNN to accelerate the learning process and enhance the computation offloading performance. If the interaction time is long enough the hotbooting Q-learning and Q-learning scheme can also converge to the optimal theoretical result in Theorem \ref{t111}. Besides, the hotbooting Q-learning based offloading scheme has less computation complexity than the fast DQN. For example, the hotbooting Q-learning strategy takes less than 4 percent of the time to choose the offloading policy in a time slot compared with the fast DQN-based scheme.

\begin{figure}[!t]
\begin{center}
\includegraphics[height= 2.0 in]{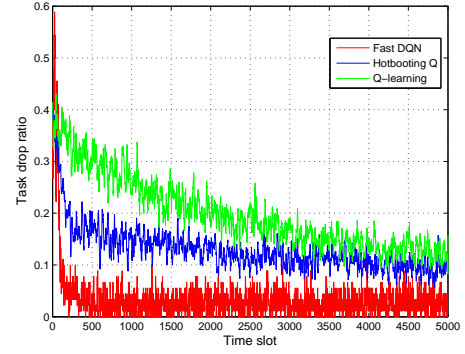} \\
\caption{Task drop ratio of the IoT computation offloading for given RF transmit power $\eta$, with the size of the computation task $C = 100$ bits.} \label {Averageperformance_Pw}
\end{center}
\end{figure}

We define the task drop ratio as the probability that a computation task fails.
The impact of the RF transmission power $\eta$ of the energy transmitter in a wireless powered MEC network is shown in Fig. \ref {Averageperformance_Pw}. The task drop ratio decreases with the RF transmit power of the energy transmitter. For instance, if the RF transmit power changes from 6 to 10, the task drop ratio of the hotbooting Q-learning based offloading scheme decreases by 82 percent. In the dynamic mobile offloading process with the RF transmit power 8, the task drop ratio of the fast DQN-based computation offloading is 53 percent less than that of hotbooting Q-learning, which is 68 percent lower than the benchmark Q-learning based offloading.

\begin{figure*}[!t]
\centering
\subfigure[]{
\begin{minipage}{5.5cm}
\centering
\includegraphics[height=4.8cm,width=6.0cm]{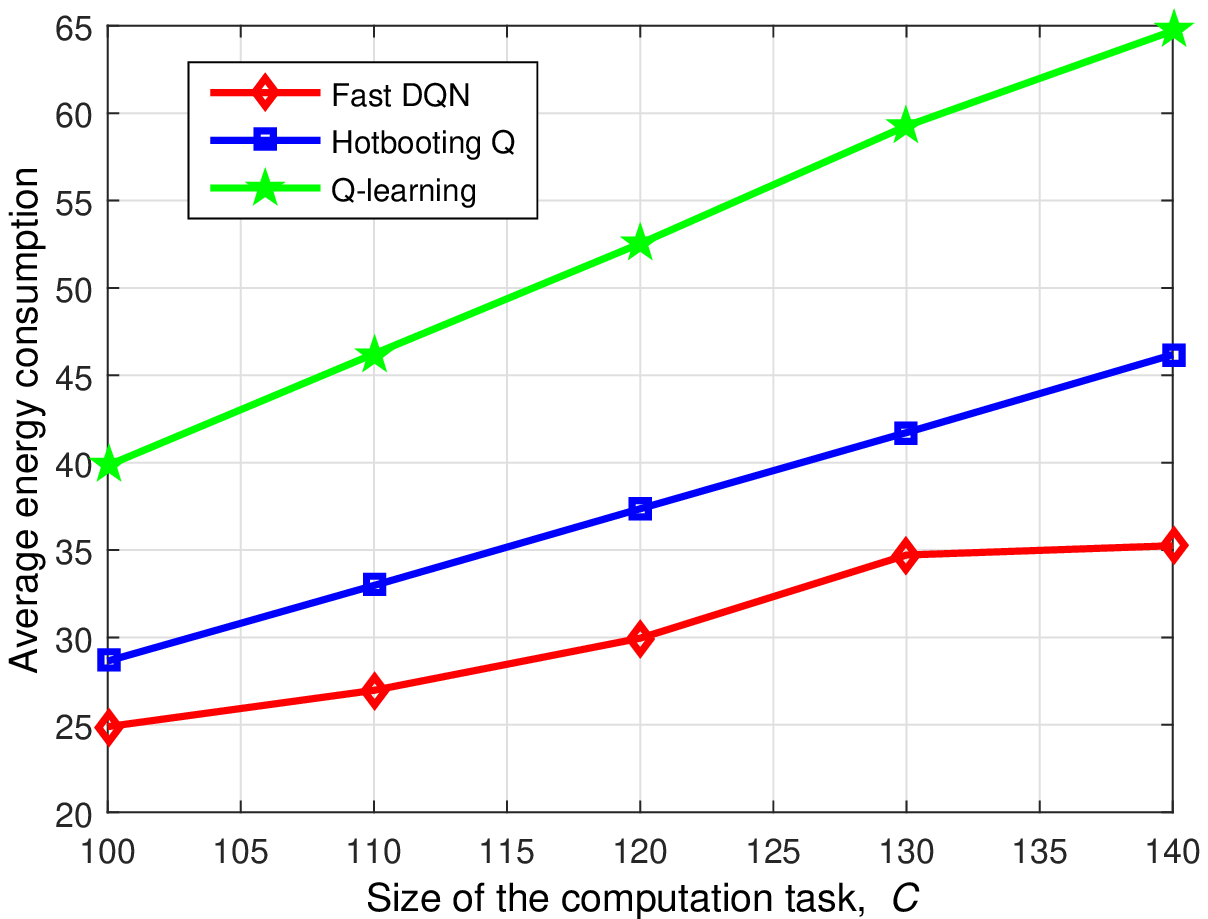}
\end{minipage}
}
\subfigure[]{
\begin{minipage}{5.5cm}
\centering
\includegraphics[height=4.8cm,width=6.0cm]{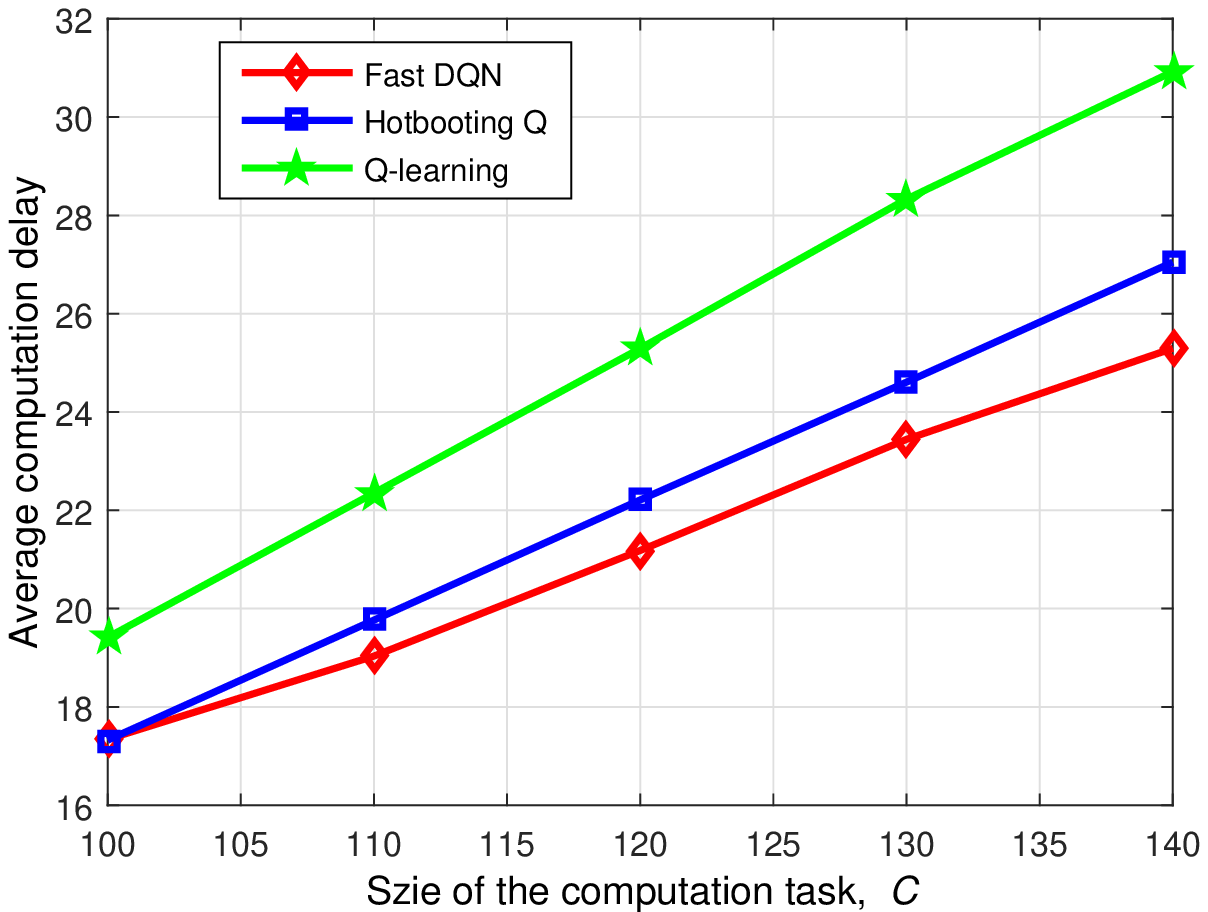}
\end{minipage}
}
\subfigure[]{
\begin{minipage}{5.5cm}
\centering
\includegraphics[height=4.8cm,width=6.0cm]{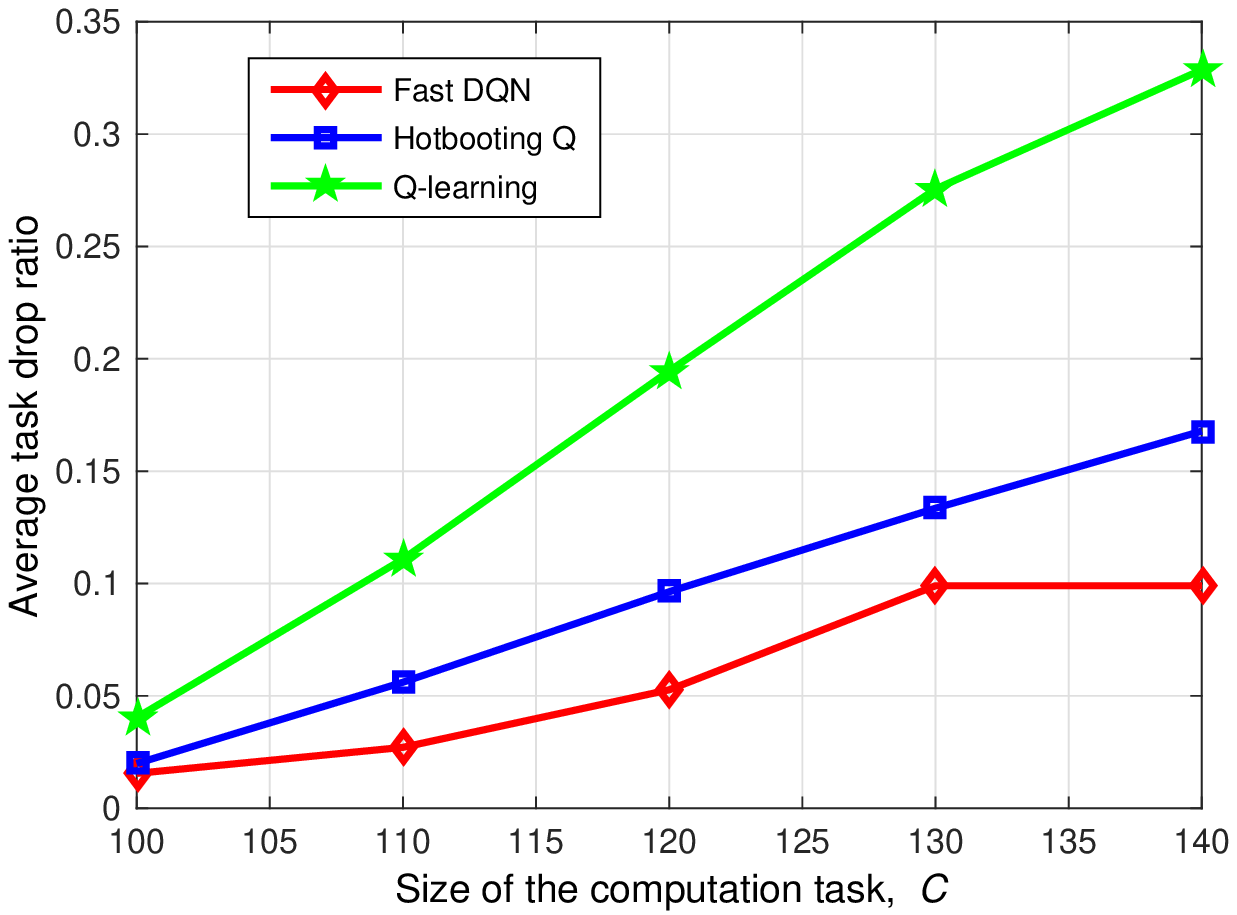}
\end{minipage}
}
\caption{Average performance of the IoT computation offloading for given size of the computation task, with the transmit power of the energy transmitter 6 $W$: a) Energy consumption; b) Computation delay; c) Task drop ratio.} \label {Averageperformance_C}
\end{figure*}

Finally, we show the relationship between the system performance and the size of the computation task $C$. As shown in Fig. \ref {Averageperformance_C}, the energy consumption, the computation delay and the task drop ratio of the IoT device increase with the size of the computation task in MEC. For instance, if the size of the computation task changes from 100 to 140, the energy consumption, the computation delay and the task drop ratio of the IoT device with fast DQN-based offloading scheme increase by 40 percent, 46 percent and 4 times, respectively. In the dynamic game with 120 bits of computation tasks, the fast DQN-based offloading scheme exceeds the hotbooting Q with 23 percent lower energy consumption, 4 percent shorter computation delay, and 50 percent lower task drop ratio.

\section*{Conclusion}
Energy harvesting MEC has become a promising technique to improve computation capabilities for self-sustainable IoT devices. In this article, we have presented the RL-based computation offloading framework for IoT devices with energy harvesting to achieve the optimal offloading policy without being aware of the MEC model and the local computation and energy consumption model. A hotbooting Q-learning based offloading scheme has been proposed for IoT devices with low complexity, which utilizes the transfer learning technique to accelerate the learning speed. We have also presented a fast DQN-based offloading scheme that uses CNN to compress the state space in the learning process. We have provided the conditions for both fully offloading and locally processing, and provided the performance bounds of the proposed RL-based offloading after convergence under two typical scenarios. Simulations have been performed for IoT devices with RF signal based wireless power transfer to verify the theoretic results and demonstrate the efficacy of the proposed RL-based schemes. In the future work, we will evaluate the performance by running real applications and show the applicability of our theoretical framework. In addition, we will consider the joint local power control and offloading decision making problem, which will be very interesting and technically challenging.

\bibliography{reference}

\begin{thebibliography}{10}

\bibitem{mao2016dynamic}
Y.~Mao, J.~Zhang, and K.~B. Letaief, ``Dynamic computation offloading for
  mobile-edge computing with energy harvesting devices,'' {\em IEEE JSAC.},
  vol.~34, no.~12, Sep. 2016, pp. 3590--3605.

\bibitem{mao2017survey}
Y.~Mao, C.~You, J.~Zhang, K.~Huang, and K.~B. Letaief, ``A survey on mobile
  edge computing: The communication perspective,'' {\em IEEE Commun. Surveys
  and Tutorials}, vol.~19, no.~4, Aug. 2017, pp. 2322-2358.

\bibitem{ulukus2015energy}
S.~Ulukus, A.~Yener, E.~Erkip, O.~Simeone, M.~Zorzi, P.~Grover, and K.~Huang,
  ``Energy harvesting wireless communications: A review of recent advances,''
  {\em IEEE JSAC.}, vol.~33, no.~3, Jan. 2015, pp. 360--381.

\bibitem{You7762913}
C.~You, K.~Huang, H.~Chae, and B.~H. Kim, ``Energy-efficient resource
  allocation for mobile-edge computation offloading,'' {\em IEEE Trans.
  Wireless Commun.}, vol.~16, no.~3, Mar. 2017, pp. 1397-1411.

\bibitem{bi2017computation}
S.~Bi, Y.~Jun, {\em et~al.}, ``Computation rate maximization for wireless
  powered mobile-edge computing with binary computation offloading,'' {\em
  arXiv preprint arXiv:1708.08810}, Aug. 2017.

\bibitem{wang2017joint}
F.~Wang, J.~Xu, X.~Wang, and S.~Cui, ``Joint offloading and computing
  optimization in wireless powered mobile-edge computing systems,'' {\em arXiv
  preprint arXiv:1702.00606}, May 2017.

\bibitem{xu2017online}
J.~Xu, L.~Chen, and S.~Ren, ``Online learning for offloading and autoscaling in
  energy harvesting mobile edge computing,'' {\em arXiv preprint
  arXiv:1703.06060}, Mar. 2017.

\bibitem{Qintroduction}
R.~S. Sutton and A.~G. Barto, ``Reinforcement learning: An introduction,'' {\em
  MIT press, Cambridge, MA}, 1998.

\bibitem{pan2010survey}
S.~J. Pan and Q.~Yang, ``A survey on transfer learning,'' {\em IEEE Trans.
  Knowl. Data Eng.}, vol.~22, no.~10, pp.~1345--1359, Oct. 2010.

\bibitem{mnih2015human}
V.~Mnih, K.~Kavukcuoglu, D.~Silver, A.~A. Rusu, J.~Veness, {\em et~al.},
  ``Human-level control through deep reinforcement learning,'' {\em Nature},
  vol.~518, no.~7540, Jan. 2015, pp. 529--533.

\bibitem{Zhang7081084}
S.~Bi, C.~K. Ho, and R.~Zhang, ``Wireless powered communication: Opportunities
  and challenges,'' {\em IEEE Commun. Mag.}, vol.~53, no.~4, Apr. 2015, pp.
  117-125.

\bibitem{Chen7307234}
X.~Chen, L.~Jiao, W.~Li, and X.~Fu, ``Efficient multi-user computation
  offloading for mobile-edge cloud computing,'' {\em IEEE/ACM Trans.
  Networking}, vol.~24, no.~5, Oct.2016, pp. 2795-2808.

\bibitem{zhang2013energy}
W.~Zhang, Y.~Wen, K.~Guan, D.~Kilper, H.~Luo, and D.~O. Wu, ``Energy-optimal
  mobile cloud computing under stochastic wireless channel,'' {\em IEEE Trans.
  on Wireless Commun.}, vol.~12, no.~9, Aug. 2013, pp. 4569--4581.

\bibitem{wang2016incentivizing}
H.~Wang and J.~Huang, ``Incentivizing energy trading for interconnected
  microgrids,'' {\em IEEE Trans. Smart Grid}, Oct. 2017.

\bibitem{lu2017}
L.~Xiaozhen, X.~Dongjin, X.~Liang, and Z.~Weihua, ``Anti-jamming communication
  game for {UAV}-aided {VANET}s,'' in {\em in Proc. IEEE Global Commun. Conf.
  (GLOBECOM)}, Singapore, Dec. 2017.

\end{thebibliography}

\bibliographystyle{ieeetr}

\end{document}